\documentclass[12pt]{revtex4-2}

\usepackage{booktabs}  
\usepackage{nicefrac}  
\usepackage[utf8]{inputenc}   
\usepackage[T1]{fontenc}
\usepackage[USenglish, ngerman]{babel}   
\usepackage[autostyle=true]{csquotes}
\usepackage{graphicx}   
\usepackage{amssymb, amsmath}   
\usepackage{xcolor}
\usepackage{physics}
\usepackage{verbatim}

\usepackage{siunitx}   
\DeclareSIUnit\atomic{\textmd{a{.}u{.}}}
\DeclareSIUnit\yr{a}
\newcommand*{\etal}{\textit{et al.}}

\begin{document}
\selectlanguage{USenglish}

\title{Polarization and driving force analysis of coherent optical shear phonons in graphite}

\author{Christian Gerbig, Silvio Morgenstern, Ahmed S. Hassanien, Marlene Adrian, Arne Ungeheuer, Thomas Baumert, and Arne~Senftleben}
\email{arne.senftleben@uni-kassel.de}
\address{University of Kassel, Institute of Physics and CINSaT, Heinrich-Plett-Straße 40, D-34132 Kassel, Germany}

\begin{abstract}
Coherent optical phonons in the degenerate inter-layer shear mode of graphite launched by femtosecond laser pulses were investigated using ultrafast electron diffraction. The collective atomic motion is shown to be polarized in a direction related to the linear polarization of the incoming laser pulse. Using a fit with a generic driven-oscillator model, the lifetime of the oscillator's driving force is determined to \SI{37(30)}{\femto\second}. This is much shorter than the lifetime of excited carriers in graphite but similar to the time scale of the loss of the hot carrier's $k$-space anisotropy.  
\end{abstract}

\maketitle 

\section{Introduction}

When a short laser pulse interacts with a crystalline solid, coherent phonons can be launched. In these phonon modes, the atoms across the crystal oscillate in phase~\cite{Kutt.1992}. This leads to time-dependent modulations of material properties such as reflectivity or lattice constants, which are accessible in time-resolved experiments~\cite{Vialla.2020, Hase.2005, Ishioka.2006, Nakamura.2008}. 

2D-layered or van-der-Waals materials consist of covalently bound atoms arranged in layers. These are attached via weak van-der-Waals forces, leading to strong anisotropy of material properties and a range of different physical phenomena~\cite{Duong.2017}. In van-der-Waals materials with hexagonal symmetry, doubly-degenerate $E$ phonon modes are present. The degeneracy allows the atoms to be displaced in any direction inside the planes, where the displacement vector can be expanded in terms of two orthogonal basis vectors. In a coherent phonon, the direction is the same for all atoms. This effectively creates a polarization of the collective oscillations related to the polarization of the incident laser light via the respective Raman tensor~\cite{Miller.2019}. In addition, phonon polarization can be influenced by strain and external fields~\cite{Sonntag.2021}.

In graphite, there are two doubly degenerate, Raman-active optical phonons of $E_{2g}$ symmetry: The in-plane stretch mode with a period of \SI{21}{\femto\second} and the inter-layer shear mode around \SI{770}{\femto\second}~\cite{Ishioka.2008, Nicklow.1972, Tan.2012}. In this work, we focus on the shear mode, which is known to exhibit coherent phonons upon excitation by short infrared laser pulses~\cite{Chatelain.2014, Boschetto.2013, Ishioka.2008, Mishina.2000}. The phonon excitation mechanism has been described as primarily impulsive, although absorption dominates the light--matter interaction over Raman-like virtual transitions~\cite{Mishina.2000}.

The formation of the coherent shear phonons in graphite is a consequence of a $\pi-\pi^*$ transition into the conduction band near the K point. The excited carrier lifetime is in the picosecond to nanosecond timescale~\cite{Malvezzi.1986}. However, on the much shorter timescale of a few \SI{10}{\femto\second}, the initially anisotropic momentum distribution of the excited carriers becomes isotropic~\cite{Mishina.2000, Mittendorff.2014}. In addition, König-Otto \etal.~\cite{KonigOtto.2016} have shown that the anisotropy survives several picoseconds if the photon energy of the exciting field is below the energy of optical phonons. It can, therefore, be inferred that the decay of the conduction band anisotropy is due to carrier--phonon scattering~\cite{Malic.2012, Trushin.2015} leads to the formation of the polarized $E_{2g}$ coherent shear phonons. 

In this paper, we analyze the excitation of coherent optical phonons of the $E_{2g}$ shear mode in few-layer graphite by femtosecond laser pulses with a central photon energy of \SI{1.56}{\electronvolt}. The phonons were measured using ultrafast electron diffraction (UED), where the oscillating atoms in the crystal modulate the intensities of the observed Bragg peaks. We will fit our results with a generic driven-oscillator model that directly calculates the influence of the coherent phonons on the Bragg peaks. From this, we can deduce the lifetime of the oscillator's driving force. In addition, we link the polarization of the incident light with the polarization of the phonons, which refers to the common displacement direction of the atoms.

\section{Materials and Methods}
\begin{figure}
	\includegraphics[width=\linewidth]{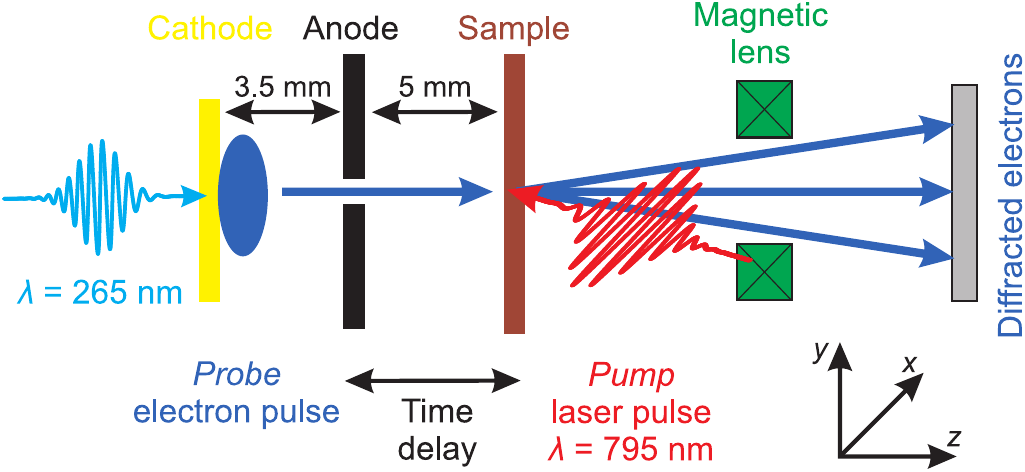}
	\caption{Illustration of the ultrafast electron diffraction set-up. Electron pulses (dark blue) are generated by UV laser pulses (light blue) at a gold-coated photocathode (yellow). The emitted electrons are accelerated to \SI{40}{\kilo\electronvolt} and pass the hole in the anode (black). Behind the sample (brown) the diffracted electrons are focused onto a position-sensitive detector (gray) by a magnetic lens (green). Time-dependent diffraction data are obtained by changing the delay between the \emph{pump} (red) and \emph{probe} pulses.}
	\label{fig:setup}
\end{figure}

The compact UED set-up is shown schematically in Figure~\ref{fig:setup}. Laser pulses with \SI{27}{\femto\second} FWHM duration from a titanium-sapphire amplifier with a central wavelength of \SI{795}{\nano\metre} at a repetition rate of \SI{1}{\kilo\hertz} were divided into two paths by a beam splitter. One beam path guided the \textit{pump} pulses via an adjustable delay path into an ultra-high vacuum chamber where they are reflected onto the graphite sample. On the second path \textit{probe} pulses are first converted to UV light at \SI{267}{\nano\metre} wavelength by third harmonic generation. The UV pulses were focused onto a gold photocathode where \num{1500(50)} electrons are liberated into the vacuum chamber. The resulting electron pulses were accelerated to \SI{40}{\kilo\electronvolt} and hit the sample only \SI{8.5}{\milli\metre} away from the cathode. Behind the sample, the diffracted electrons were focused onto a position-sensitive detector by a magnetic lens. Unscattered electrons were prevented from hitting the detector by a beam stop. Time-dependent data was obtained by changing the delay between the \textit{pump} and \textit{probe} pulses. More details on the set-up and characterization of the instrumental resolution have been reported by Gerbig~\etal.~\cite{Gerbig.2015}. The sample is placed such that the graphite planes with crystal axes $a$ and $b$ are in the $(x,y)$ plane of our laboratory frame while the electron pulses propagate in $+z$ direction, i.e. parallel to the $c$ axis of the crystal. The \textit{pump} pulses propagate approximately in $-z$ direction and are linearly polarized along the $x$ axis.

During the present work, the mean fluence of the \textit{pump} pulse in the area of the sample \textit{probed} by the electrons is \SI{6.2(4)}{\milli\joule\per\centi\metre\squared}. The RMS temporal resolution of the \textit{pump--probe} experiment under the current conditions is \SI{101(14)}{\femto\second} while the transverse coherence length of the electron pulses on the sample is \SI{3.8(6)}{\nano\metre}. More than 100 Bragg diffraction peaks with reciprocal lattice vectors ranging up to \SI{18.4}{\per\angstrom} were recorded. The time delay $t$ was scanned in \SI{50}{\femto\second} steps from \SI{-1.7}{\pico\second} (\textit{probe} before \textit{pump}) to \SI{10}{\pico\second} (\textit{probe} after \textit{pump}).

Our free-standing sample was exfoliated mechanically~\cite{Meyer.2008} from a macroscopic single crystal of natural graphite and later placed on a Quantifoil 200 mesh TEM grid with a line spacing of \SI{90}{\micro\metre}. The sample flake covered little more than one grid cell, closely matching the electron pulse diameter of $\approx\SI{100}{\micro\meter}$. The thickness of the flake was determined by reflective contrast while resting on an oxide-coated silicon wafer~\cite{Skulason.2010} to 9 layers with a confidence of at least \SI{96}{\percent}.


\section{Results}

\begin{figure}
	\includegraphics[width=\linewidth]{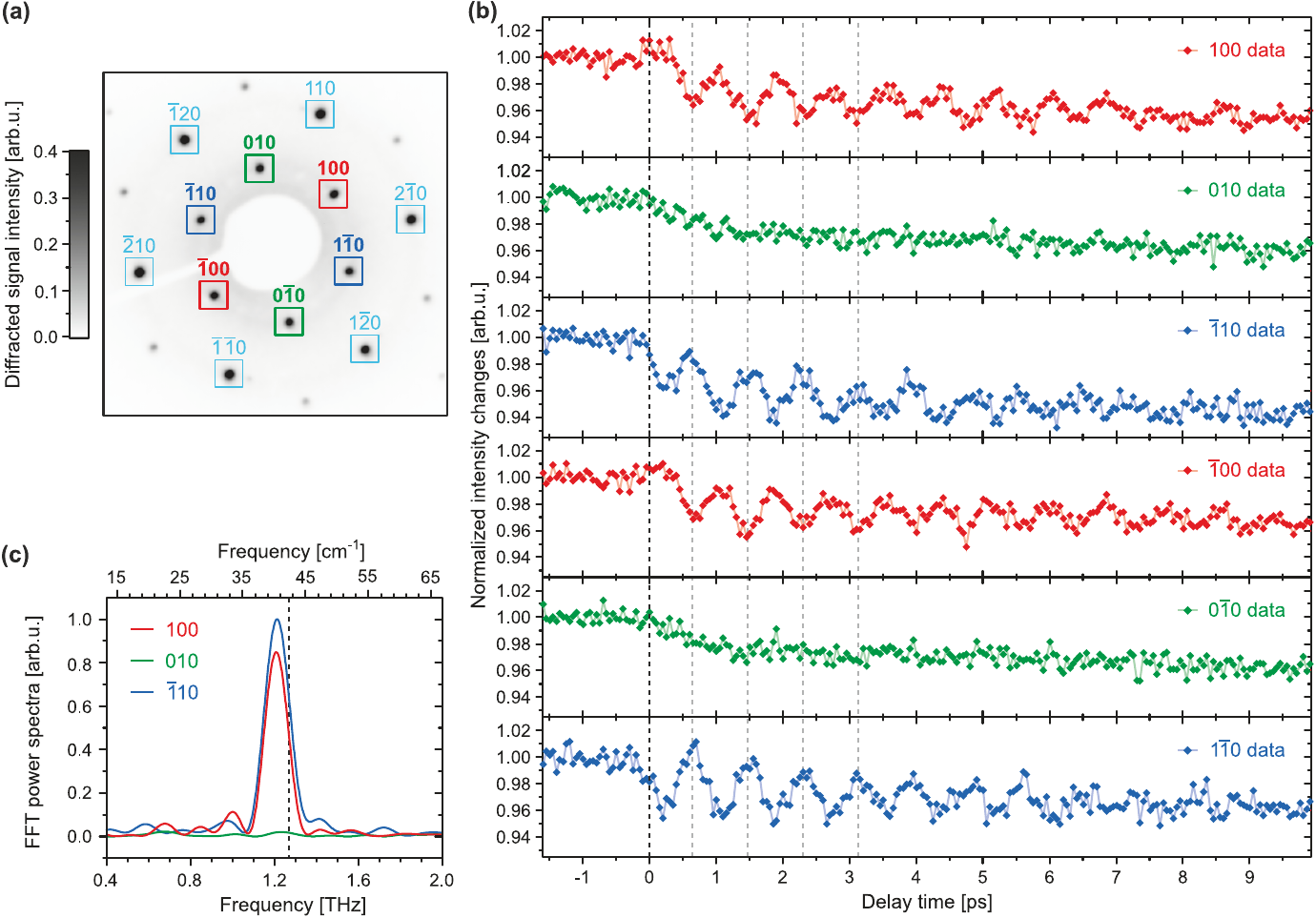}
	\caption{(a) Part of the diffraction pattern of graphite showing the \{$1\,0\,0$\} and the \{$1\,1\,0$\} Bragg peaks. (b) Time-dependent integral intensity of the \{$1\,0\,0$\} reflexes for the first \SI{10}{\pico\second} after excitation, which happens at zero delay time. (c) Fourier transformation spectra of the traces shown in (b). The dashed vertical line indicates the shear mode frequency measured for 9-layer graphite by Boschetto~\etal.~\cite{Boschetto.2013, Ferrari.2013}.}
	\label{fig:resultsRaw}
\end{figure}

Figure~\ref{fig:resultsRaw}~(a) displays the central part of the diffraction pattern of our nine-layer graphite sample. The occurrence of clear Bragg peaks mirrors the monocrystallinity of the flake. This work focuses on the evolution of the integral Bragg peak intensities during the first \SI{10}{\pico\second} after excitation. For each time delay $t$, the intensities $I(h\,k\,l,t)$ were extracted by fitting two-dimensional Gaussian distributions to the Bragg peaks. At large negative time delays, the \textit{probe} electrons pulse precedes the \textit{pump} pulse and diffracts on the undisturbed lattice, yielding the constant Bragg peak intensities $I(h\,k\,l,-\infty)$. To visualize the change in lattice structure induced by the \textit{pump} pulse we introduce the relative intensity $\mathcal{I}(h\,k\,l,t)=I(h\,k\,l,t)/I(h\,k\,l,-\infty)$ where $I(h\,k\,l,-\infty)$ is determined by averaging the Bragg peak intensities for negative delay values. The measured $\mathcal{I}(h\,k\,l,t)$ for the first order reflexes are presented in Figure~\ref{fig:resultsRaw}~(b).

After the excitation at zero time delay, the intensities show two developments: Firstly, the intensity of the peaks drops. This decrease is observed in all recorded Bragg reflexes and has been attributed to the heated lattice's Debye--Waller effect~\cite{Gerbig.2015}. Additionally, the traces for the $1\,0\,0$, $\bar{1}\,1\,0$, $\bar{1}\,0\,0$ and $0\,\bar{1}\,0$ Bragg peaks show pronounced oscillations with fading amplitudes which we attribute to coherent optical phonons of the $E_{2g}$ shear mode. One can see that Bragg peaks with opposite directions (compare Figure~\ref{fig:resultsRaw}~(a)) show the same temporal behavior per Friedel's law. In the following, we will sum up the time-dependent signals for pairs of opposite peaks and refer to these combined data as $1\,0\,0$, $0\,1\,0$ and $\bar{1}\,1\,0$, respectively. 

A precise knowledge of time zero is crucial for quantitative analysis of the signals. We understand this time as the centroid of the cross-correlation between the excitation laser and electron pulse at the sample. Here, we determine it from the fast decrease of diffraction intensity for Bragg peaks that do not exhibit oscillating temporal behavior. 65 traces were fitted with bi-exponential decays according to Schäfer \etal.~\cite{Schafer.2011} additionally convoluted with the electron pulse duration. From these, time zero was found with an rms accuracy of \SI{26}{\femto\second}. All temporal traces shown in this work use the time zero determined by these bi-exponential fits.

Fourier transformation (Figure~\ref{fig:resultsRaw}~(c)) of the temporal traces exhibit a central oscillation period of \SI{826(14)}{\femto\second} for all reflexes. This is considerably longer than the reported values of \SI{790}{\femto\second} for the inter-layer shear phonon in 9-layer graphite by Raman scattering~\cite{Tan.2012, Boschetto.2013}. One reason for this frequency redshift could be the relatively high fluence of our \textit{pump} pulse. However, for bulk graphite, significantly smaller frequency shifts for the shear mode have been reported at smaller \cite{Mishina.2000} and much larger fluences \cite{Lu.2005}. But we also measured coherent shear phonons in a \SI{42(3)}{\nano\metre} thick graphite nanofilm at a fluence of \SI{1.8}{\milli\joule\per\centi\metre\squared}. There, we observed an oscillation period of \SI{778(5)}{\femto\second}, slightly longer than the \SI{767}{\femto\second} observed in Raman scattering. These findings indicate that a fluence dependence could arise for very thin samples. Another option in our 9-layer sample could be the population of lower-frequency branches of the shear mode as observed in various 2D-layered materials~\cite{Zhao.2013, Zhang.2013b, Liang.2017}. However, this seems unlikely because then one would expect an asymmetric frequency spectrum, which we do not observe.


\begin{figure}
	\includegraphics[width=\linewidth]{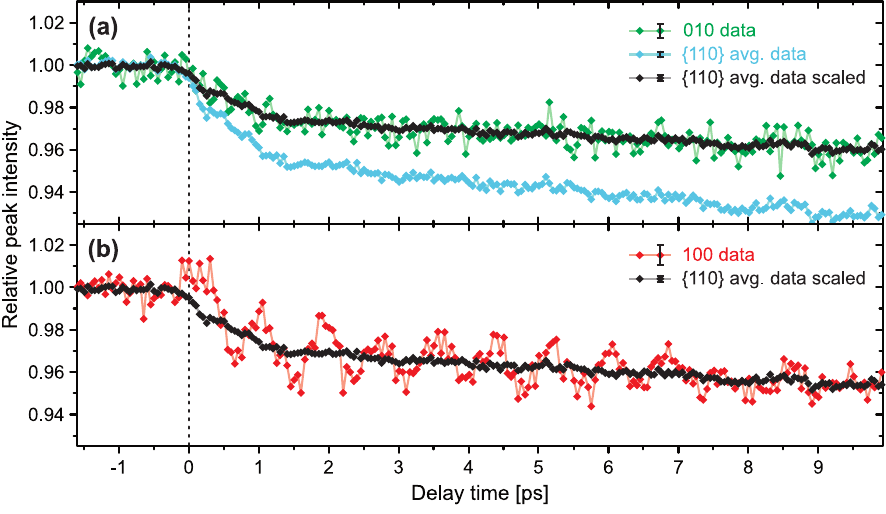}
	\caption{Procedure for isolation of coherent phonon signals: (a) Time-dependent relative intensity for the $0\,1\,0$ Bragg peak (green) as well as averaged over the $\{1\,1\,0\}$ Bragg peaks unscaled (blue) and scaled to minimize the mean quadratic deviation from the $0\,1\,0$ trace (black). (b) Scaled black trace from~(a) overlaid with time-dependent relative intensity of the $1\,0\,0$ Bragg peak (red).}
	\label{fig:backgrundSubtraction}
\end{figure}

The relative time-dependent Bragg peak intensities $\mathcal{I}(h\,k\,l,t)$ as displayed in Figure~\ref{fig:resultsRaw}~(b) are products of the Debye--Waller contribution $\mathcal{I}_{DW}(h\,k\,l,t)$ and the intensity modulation due to the coherent phonons $\mathcal{I}_{CP}(h\,k\,l,t)$. To isolate the latter, we divide the traces in Figure~\ref{fig:resultsRaw}~(b) by their respective Debye--Waller contributions $\mathcal{I}_{DW}(h\,k\,l,t)$. Therefore, we make use of the signal from the $\{1\,1\,0\}$ Bragg peaks which do not exhibit any coherent-phonon-related oscillations therefore purely represents $\mathcal{I}_{DW}(\{1\,1\,0\},t)$. As presented in Figure~\ref{fig:backgrundSubtraction}~(a), the Debye--Waller contribution to the first order Bragg peaks, $\mathcal{I}_{DW}(\{1\,0\,0\},t)$ is approximated by scaling $\mathcal{I}_{DW}(\{1\,1\,0\},t)$ to minimize the mean square deviation from $\mathcal{I}(0\,1\,0,t)$. Figure~\ref{fig:backgrundSubtraction}~(b) demonstrates that the scaled $\{1\,1\,0\}$ signal matches the baseline of the shear mode oscillations. Subsequently, the $1\,0\,0$, $0\,1\,0$ and $\bar{1}\,1\,0$ time-dependent intensities are divided by $\mathcal{I}_{DW}(\{1\,0\,0\},t)$. Consequently, the isolated traces (see Figure~\ref{fig:resultsFitted}) contain only the oscillatory relative intensities $\mathcal{I}_{CP}(h\,k\,l,t)$ due to the coherent shear phonons. 

The first maximum or minimum of the oscillations is almost one-quarter of the period away from time zero. This is a signature of impulsive phonon excitation~\cite{Matsumoto.2014, Ishioka.2010, Misochko.2004}. In the following section, we will develop a model to fit the experimental results directly. The fit allows the determination of the lifetime $\tau$ of the conduction band anisotropy, while the initial phase of the oscillation is no explicit parameter.

\section{Discussion}

\begin{figure}
	\includegraphics[width=.8\linewidth]{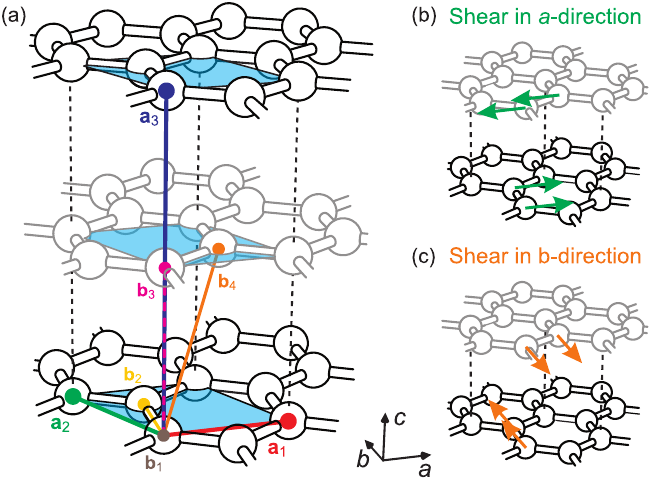}
	\caption{(a) Basis vectors used to describe the unit cell of graphite. $\vectorbold{a}_1$, $\vectorbold{a}_2$ and $\vectorbold{a}_3$ are lattice vectors, while $\vectorbold{b}_1$, $\vectorbold{b}_2$, $\vectorbold{b}_3$ and $\vectorbold{b}_4$ give the location of the atoms in the unit cell. (b, c) Basis vectors for the shear displacement $\mathcal{P}_a, \mathcal{P}_b$ along the crystal's $a$ and $b$ directions, respectively.}
	\label{fig:basisVectors}
\end{figure}

In order to understand the excitation mechanism of the observed shear mode and the angular anisotropy, we are going to fit the measured Bragg peak intensity modulations $\mathcal{I}_{CP}(\{1\,0\,0\},t)$ with a model comprising a generic driven oscillator for the atomic displacements as well as the calculation of the time-dependent Bragg peak intensities from these displacements. Note that this model does not make any assumptions on the material except for the equilibrium positions of the atoms and that we, therefore, do not use material-specific parameters.

First, we will deduce how the coherent optical phonons influence the Bragg peak intensities. Therefore, we use the four-atom unit cell for graphite illustrated in Figure~\ref{fig:basisVectors}. As the crystal planes follow a $AB$ stacking order, two planes must be included, each contributing two atoms to the unit cell. We define the covalently bound planes of graphite to be in the $(a,b)$ plane of the crystal-centered coordinate system, where the $a$ axis is parallel to the $\vectorbold{a}_1$ unit vector (compare Figure~\ref{fig:basisVectors}~(a)).

The shear phonon is a transverse optical phonon with its wave vector along the $c$-axis that only slightly disperses between $\Gamma$ and the $A$ point~\cite{Nicklow.1972, Wirtz.2004}. Being of $E_{2g}$ symmetry, it has two degenerate displacement directions in the $(a,b)$ plane. In other words, the coherent phonon oscillations can be \emph{polarized} along any direction inside the basal plane, and the relative atomic displacement can always be decomposed in the two basis components $\mathcal{P}(a)$ and $\mathcal{P}(b)$ along the $a$ and $b$ axes, respectively (compare Figure~\ref{fig:basisVectors}~(b) and~(c)).

Under the influence of the coherent shear phonons, we assume that the $A$ plane with the atoms located at $\vectorbold{b}_1$ and $\vectorbold{b}_2$ stays at rest while the $B$ plane with atoms at $\vectorbold{b}_3$ and $\vectorbold{b}_4$ is shifted in the $a$ and $b$ directions by the relative displacements $\mathcal{P}_a(t)$ and $\mathcal{P}_b(t)$, respectively. This results in the modified atomic basis vectors $\vectorbold{\widetilde{b}}_i(t)$:
\begin{equation}
	\begin{array}{ll}
		\vectorbold{\widetilde{b}}_1(t)=(0,0,0);&\vectorbold{\widetilde{b}}_2=\qty(0,\frac{d_a}{\sqrt{3}},0)\\
		\multicolumn{2}{l}{\vectorbold{\widetilde{b}}_3(t)=\qty(0,0,\frac{d_c}{2})+d_a\qty(\mathcal{P}_a,\mathcal{P}_b,0) }\\
		\multicolumn{2}{l}{\vectorbold{\widetilde{b}}_4(t)=\qty(\frac{d_a}{2},\frac{d_a}{2\sqrt{3}},\frac{d_c}{2})+d_a\qty(\mathcal{P}_a,\mathcal{P}_b,0) }
	\end{array}
	\label{eq:basisVectors}
\end{equation}
where $d_a=\SI{2.461}{\angstrom}$ and $d_c=\SI{6.709}{\angstrom}$ are the lattice constants of graphite~\cite{Martienssen.2005}. The time-dependent relative Bragg peak intensity $\mathcal{I}(h\,k\,l,t)$ is given as
\begin{equation}
	\mathcal{I}(h\,k\,l,t)=\frac{I(h\,k\,l,t)}{I(h\,k\,l,-\infty)}=\mathcal{I}_{DW}(h\,k\,l,t)\frac{\qty|\sum_{n=1}^4e^{-i\vectorbold{G}_{h\,k\,l}\cdot\vectorbold{\widetilde{b}}_n(t)} |^2}{\qty|\sum_{n=1}^4e^{-i\vectorbold{G}_{h\,k\,l}\cdot\vectorbold{b}_n} |^2}
	\label{eq:calculateRelativeBragg}
\end{equation}
where $\vectorbold{G}_{h\,k\,l}$ is the corresponding reciprocal lattice vector. The fractional expression at the end of eq.~\ref{eq:calculateRelativeBragg} equals the coherent phonon induced Bragg peak intensity modulations $\mathcal{I}_{CP}(h\,k\,l,t)$, which for displacements much smaller than $a$ calculates for the three relevant peaks to
\begin{equation}
	\begin{array}{l}
		\mathcal{I}_{CP}\qty(100,t)\approx 1+2\pi\qty(\sqrt{3}\,\mathcal{P}_a(t)+\mathcal{P}_b(t))\\
		\mathcal{I}_{CP}\qty(010,t)\approx 1-4\pi\,\mathcal{P}_b(t)\\
		\mathcal{I}_{CP}\qty(\bar{1}\,1\,0,t)\approx 1+2\pi\qty(-\sqrt{3}\,\mathcal{P}_a(t)+\mathcal{P}_b(t)).
	\end{array}
	\label{eq:braggDisplacementsDerivation}
\end{equation}
We should add that in the same approximation, the shear mode induced relative intensities for the \{$1\,1\,0$\} Bragg peaks vanish, which is in agreement with our observations and justifies the use of those peaks to isolate the Debye--Waller and coherent-phonon contributions in the ${100}$ Bragg peaks.

\begin{figure}
	\includegraphics[width=\linewidth]{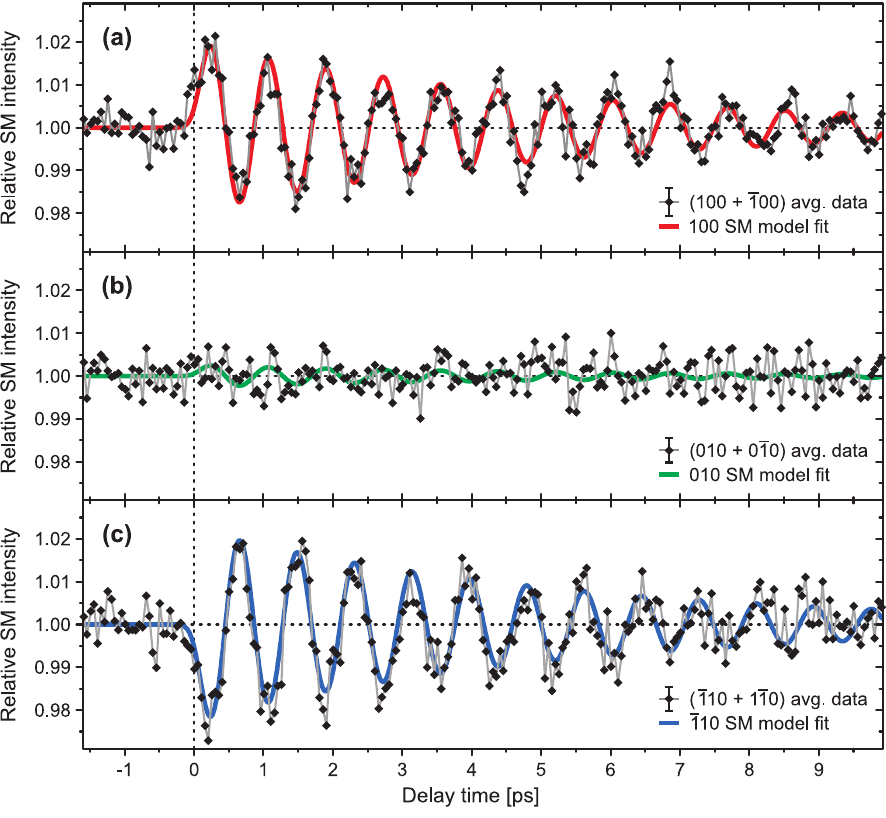}
	\caption{Time-dependent relative shear mode (SM) diffraction intensities for the (a) $1\,0\,0$, (b) $0\,1\,0$, and (c) $\bar{1}\,1\,0$ reflections together with fits from the model discussed in the text.}
	\label{fig:resultsFitted}
\end{figure}

Now we assign the relative lattice displacements $\mathcal{P}_{a,b}(t)$ to coherent phonons induced by the photo-excitation of graphite. In the most general form, the displacements must fulfill the differential equations of a damped harmonic oscillator driven by the force $F(t)$

\begin{equation}
	\mu \left[ \frac{\text{d}^2\mathcal{P}_{a,b}(t)}{\text{d} t^2}+\frac{2}{\beta}\frac{\text{d} \mathcal{P}_{a,b}(t)}{\text{d} t} + \qty(\frac{2\pi}{T_0})^2\mathcal{P}_{a,b}(t) \right]=F(t)\Pi_{a,b}(\theta)
	\label{eq:drivenOscillator}
\end{equation}
with the reduced lattice mass $\mu$, the damping time $\beta$ and the natural (undamped) period of oscillation $T_0$. 

$\Pi_{a,b}(\theta)$ are the components of a unit vector that determines the polarization of the driven phonons in the $(a,b)$ plane and depends on the angle $\theta$ between the crystal's $a$ axis and the laboratory-frame $x$ axis along which the laser is polarized:
\begin{equation}
	\Pi_{a} = \vectorbold{E} \mathcal{R}_{a}(\theta) \vectorbold{E}=\sin(-2\theta),\;\Pi_{b} = \vectorbold{E} \mathcal{R}_{b}(\theta) \vectorbold{E}=\cos(2\theta)
	\label{eq:anisotropyPi}
\end{equation}

where $\mathcal{R}_{a,b}(\theta)$ are Raman tensors for $E_{2g}$ phonons~\cite{Loudon.1964} that have been rotated to encompass the orientation of the crystal and observation direction of the detector. From equation~\ref{eq:anisotropyPi}, it follows that the angle between the polarization of the coherent phonon and the crystal's $a$-axis amounts to $\ang{90}+2\theta$.

In an opaque material such as graphite, light absorption leads to the promotion of charge carriers to an excited band. This changes the lattice potential. Consequently, the lattice is driven by a force proportional to the density of excited carriers~\cite{Cheng.1991, Zeiger.1992}. However, due to the anisotropic nature of $E_{2g}$ phonons, we need to consider excited carriers that display a $k$-space anisotropy~\cite{Riffe.2007} which is the case for the $\pi$--$\pi^*$ excitation in graphite or graphene~\cite{Mishina.2000, Malic.2011}. Including an exponential lifetime $\tau$ of the excited carrier anisotropy, the time-dependent force becomes
\begin{equation}
	F(t) = C e^{-t/\tau}\int_{-\infty}^t \abs{\vb{E}(t')}^2e^{t/\tau}dt'
	\label{eq:DrivingForce}
\end{equation}
with a constant $C$ and the intensity profile $\abs{\vb{E}(t)}^2$ of the exciting laser pulse, which is approximated by a Gaussian profile with a full width at half maximum of \SI{27}{\femto\second}. Note that equation~\ref{eq:DrivingForce} neglects transitions to virtual states as would be appropriate for situations where both resonant excitation and Raman transitions are possible~\cite{Riffe.2007}. However, in graphite at our incident wavelength, this simplification is justified, as the derivative of the real part of the dielectric function is much smaller than the imaginary part of the dielectric function~\cite{Ahuja.1997,Kuzmenko.2008}. Additionally, Mishina \etal.~\cite{Mishina.2000} identified real excitation as the main source to induced shear phonons in graphite while at the same time observing a Raman-like anisotropy.


With the information of equations~\ref{eq:DrivingForce} and~\ref{eq:anisotropyPi}, numerical solutions to equation~\ref{eq:drivenOscillator} were calculated. The resulting lattice displacements $\mathcal{P}_{a,b}(t)$ were convoluted with the temporal intensity profile of the electron pulses and inserted into equation~\ref{eq:braggDisplacementsDerivation} to obtain the coherent-phonon-induced relative Bragg peak intensities $\mathcal{I}_{CP}(h\,k\,l,t)$. These were fitted to the experimental traces simultaneously as displayed in Figure~\ref{fig:resultsFitted} by iterating the numerical solutions to equation~\ref{eq:drivenOscillator} varying the following parameters: crystal orientation $\theta$, excitation lifetime $\tau$, period of oscillation $T_0$, damping time $\beta$, and the intensity constant $C$. As seen in Figure~\ref{fig:resultsFitted}, the model fits the measured data nicely. 

\begin{figure}
	\includegraphics[width=\linewidth]{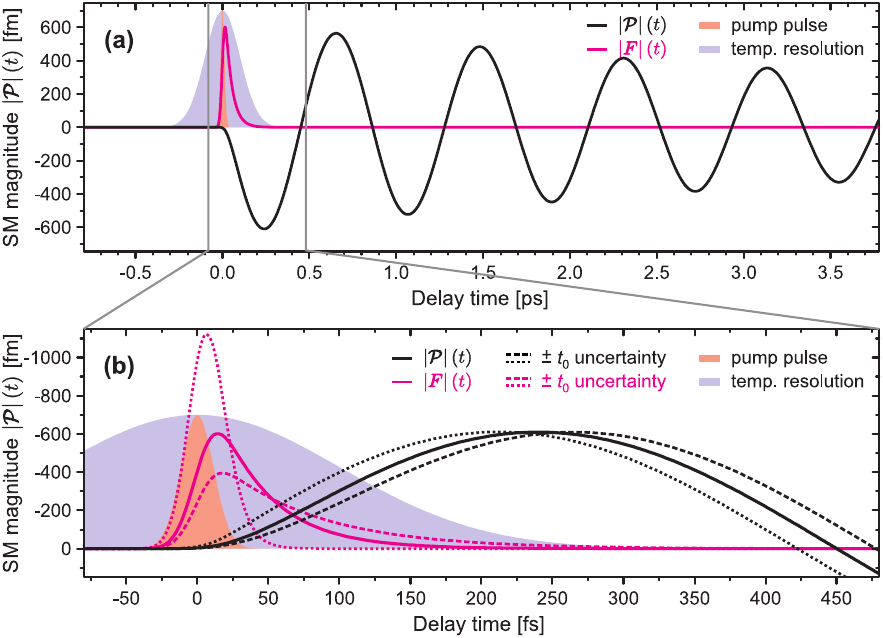}
	\caption{Shear lattice displacement $|\mathcal{P}(t)|=\sqrt{{\mathcal{P}_a(t)}^2+{\mathcal{P}_b(t)}^2}$ (black lines) as fitted to the experimental data with the model described in the text. From the same fit results, the driving force $F(t)$ (red lines). For comparison, the estimated temporal duration of the \emph{pump} and \emph{probe} pulses are shown as orange and blue shades, respectively. Panel (b) displays a smaller range of the delay axis and, in addition, contains error margins of the fits as dashed and dotted lines. Note that the phonon displacement has been mirrored in panel~(b).}
	\label{fig:ResultsTimeZero}
\end{figure}

In order to account for the uncertainty of time zero (see above), we perform the model fits for time axes shifted by $\pm\SI{26}{\femto\second}$, which is illustrated in Figure~\ref{fig:ResultsTimeZero}. Panel (a) compares the timescales of the coherent phonons with the extent of the \textit{pump} pulse, the temporal resolution of the UED set-up and the driving force $F(t)$ obtained at the optimal time zero. Panel (b) shows a close-up with the $F(t)$ and $|\mathcal{P}(t)|=\sqrt{{\mathcal{P}_a(t)}^2+{\mathcal{P}_b(t)}^2}$ additionally obtained for time zero shifted by its uncertainty in positive (dashed lines) and negative (dotted lines) directions. From these three cases, we obtain the final value for the driving force's time constant $\tau=\SI{37(30)}{\femto\second}$. This time corresponds to the decay of $k$-space anisotropy of carrier excitation in graphite~\cite{Mishina.2000, Gruneis.2003}, which is caused by carrier--phonon scattering~\cite{Malic.2012}. Our time scale corresponds well with theoretical predictions and optical measurements of this loss in excited-carrier anisotropy~\cite{Malic.2012, Mittendorff.2014}. 

The oscillation period $T_0=\SI{827(1)}{\femto\second}$ agrees with the value obtained from Fourier transformation and is much larger than the lifetime $\tau$, which leads to the observed impulsive nature of coherent phonon excitation although absorption is dominant. To quantify the impulsive behaviour, we fit a damped cosine function to the solution of the lattice displacement $|\mathcal{P}(t)|$. From this, we estimate the initial phase of the coherent phonon oscillation to $\phi\approx\ang{74}$. The same value is obtained when calculating the phase following a simple model that only considers the excitation of the crystal to another band~\cite{Zeiger.1992} with vanishing damping. The agreement of this calculated phase with the fitted one confirms that the contribution of virtual transitions is negligible and that it is not necessary to use a more elaborate model~\cite{Riffe.2007} to calculate the experimental phase correctly. Our system does not reach the ideal impulsive limit of $\phi=\ang{90}$ because the lifetime of the anisotropy in the conduction band is finite.

It should be noted that we do not observe any delay in the rise of coherent phonon population, as was detected with time-resolved Raman spectroscopy~\cite{Yang.2017} for the higher energy $G$-mode phonon, which is another $E_{2g}$ mode at the $\Gamma$ point that corresponds to in-plane vibrations. They reported a delay of \SI{65}{\femto\second}, which has been attributed to a sequential coupling process, where the hot carriers first excite an $A_1'$ in-plane mode at the $K$ point, which then couples to the in-plane $E_{2g}$ mode. We derive a direct coupling of the $k$-space carrier anisotropy to the $E_{2g}$ shear mode.

Furthermore, from our fit, we can calculate the maximum shear mode displacements to $\SI{609(128)}{\femto\metre}$, which is more than a factor of \num{400} smaller than the in-plane lattice constant $a$. Consequently, the assumption of small displacements in the derivation of equation~\ref{eq:braggDisplacementsDerivation} is justified. 

The crystal orientation angle $\theta$ between the polarization of the laser and the $a$ axis of the crystal was fitted to \SI{-42.0(7)}{\degree}. From this value and the $\cos(2\theta)$ dependence of $\mathcal{P}_b$ (equation~\ref{eq:anisotropyPi}) as well as the calculated value of $\mathcal{I}_{CP}\qty(010,t)$ (equation~\ref{eq:braggDisplacementsDerivation}) follows the near-extinction of oscillations in the $0\,1\,0$ trace. On the other hand, the opposite phase of the oscillations between the $1\,0\,0$ trace and the $\bar{1}\,1\,0$ trace stems from the opposite signs to $\mathcal{P}_a$ in equation~\ref{eq:braggDisplacementsDerivation}. The fitted angle can be compared to the observed diffraction pattern (Figure~\ref{fig:resultsRaw}~(a)): From $\theta=\SI{-42.0(7)}{\degree}$ we derive an angle of \ang{-17.0(7)} between the laser polarization and the direction of the 100 Bragg peak. Considering the magnetic lens's rotation of the diffraction pattern, this is in good agreement.

\section{Conclusion}

We have investigated the formation of coherent inter-layer shear phonons in graphite after optical excitation with a femtosecond laser pulse. Using ultrafast electron diffraction as a probe, we could image the polarization of the collective atomic motion inside the graphite basal planes. Employing a fit with a generic driven-oscillator model, we could determine the lifetime of the oscillations' effective driving force to \SI{37(30)}{\femto\second}, which agrees with the time scale on which the $k$-space anisotropy of optically excited carriers disappears. Including Raman tensors in our model, we can link the phonon polarization to the alignment of the crystal relative to the polarization of the pump light. This knowledge is an essential prerequisite to arbitrarily designing the phonon polarization with shaped laser pulses. 

\section*{References}

\bibliographystyle{iopart-num}
\bibliography{COPgraphite}

\end{document}